\documentclass[%
 aip,
 jmp,%
 amsmath,amssymb,
 reprint,%
]{revtex4-1}

\usepackage{graphicx}
\usepackage{dcolumn}
\usepackage{bm}

\begin{document}


\title{Light capsules shaped by curvilinear meta-surfaces}

\author{Mahin Naserpour}
 \altaffiliation[Also at ]{Physics Department, College of Sciences, Shiraz University, Shiraz 71946-84795, Iran.}
\author{Carlos J. Zapata-Rodr\'{i}guez}%
 \email{carlos.zapata@uv.es}
\affiliation{Department of Optics and Optometry and Vision Science, University of Valencia, \\ Dr. Moliner 50, Burjassot 46100, Spain}

\author{Abdolnaser Zakery}
 \homepage{E-mail: zakeri@susc.ac.ir}
\affiliation{Physics Department, College of Sciences, Shiraz University, \\ Shiraz 71946-84795, Iran}

\author{Juan J. Miret}
\affiliation{Department of Optics, Pharmacology and Anatomy, University of Alicante, \\ P.O. Box 99, Alicante, Spain}

\date{\today}

\begin{abstract}
We propose a simple yet efficient method for generating in-plane hollow beams with a nearly-full circular light shell without the contribution of backward propagating waves. 
The method relies on modulating the phase in the near field of a centro-symmetric optical wavefront, such as that from a high-numerical-aperture focused wave field.
We illustrate how beam acceleration may be carried out by using an ultranarrow non-flat meta-surface formed by engineered plasmonic nanoslits. 
A mirror-symmetric, with respect to the optical axis, circular caustic surface is numerically demonstrated that can be used as an optical bottle. 
\end{abstract}

\maketitle

\section{Introduction}
\label{intro}
Optical bottle beams that contain an intensity null point generated considerable interest due to their potential application in optical confinement of laser-cooled atoms and in optical tweezers systems \cite{Ozeri99,Arlt00,McGloin03,Yelin04}. 
This bottling effect appears when plane waves in the Fourier expansion of beams interfere destructively to generate a bottle or a null intensity point at the center of the beam.
Conventional hollow beams used in optical tweezers and following the abovementioned basic concept are Laguerre-Gaussian beams. 
More recently, an optical  scheme for realizing a supersized light capsule was presented theoretically and experimentally in air,  where the volume of darkness inside the light capsule can be changed in a wide range \cite{Wan14}. In this case, a binary phase mask was used to create such a tunable antiresolution region.  

Alternatively, wave fields exhibiting a peak intensity of curvilinear trajectory that additionally constitutes the closed boundary of a geometrical shadow, instead of a null point, have been proposed to be used as optical bottles \cite{Chremmos11,Chremmos12}.
A well-known example of this kind of wave fields are the so-called Airy beams.
Unlike ordinary optical wavefronts, Airy beams transversely accelerate, preserving their intensity profile throughout propagation but the profile’s features follow a curved path \cite{Siviloglou07b}.
By appropriately superimposing mirror-symmetric Airy beams, the field amplitude oscillates outward of a dark closed region \cite{Efremidis10}.
These theoretical predictions were subsequently verified by experimental observations \cite{Papazoglou11}.
The resulting caustic surface that abruptly emerges and subsequently autofocuses was used as an optical bottle to trap and manipulate dielectric microparticles \cite{Zhang11b,Zhang13,Zhang14}.

The intensity peak of an Airy beam follows a parabolic trajectory much like the ballistics of projectiles.
The shape of the optical bottle produced by this sort of paraxial accelerating beam is also governed by the parabolic caustic curve.
This fundamental limit may be overcome since Fourier-space generation of arbitrary accelerating beams allow the propagation along controlled curvilinear trajectories \cite{Froehly11,Zhang12b,Mathis13}.
For instance, nonparaxial accelerating waves in two dimensions, whose intensity profiles are roughly preserved within a range of propagation distances, and whose maxima follow semicircular paths, were recently proposed \cite{Zhang12,Courvoisier12}. 
Such a beam with Bessel signature should be able to bend from a launch angle of $+90^\circ$ ---perpendicular to the original direction of propagation--- all the way to an angle of zero, and immediately accelerating to propagate at angles close to $-90^\circ$ \cite{Kaminer12,Zapata14c}.
The wave field will propagate backward in the remainder half circular trajectory, a fact that in practice limits the formation of light capsules of circular symmetry.

In this study we extend these ideas to synthesize an optical bottle exhibiting all-inclusive circular symmetry, integrating forward-only propagating waves.
To construct these waves we combine two concentric incomplete Bessel beams by applying the concept of symmetrization in a notably dissimilar way of that discussed in Ref.~\cite{Vaveliuk14}.
For that purpose we follow a method previously shown in Ref.~\cite{naserpour2014highly} that potentially is appropriate for using in a compact-sized system, which makes use of a non-planar meta-surface sustained by dispersion localities. 
Our multilayered metal-dielectric nanostructure is piecewise periodic in the angular coordinate, thus enabling to transform a high-aperture focused beam into a mirror-symmetric accelerating beam nearly enclosing a circular region.
\section{Theoretical preliminaries}
\label{sec:1}
Our study is based on the efficient transformation of a converging wave field into a mirror-symmetric accelerating beam with a prescribed circular trajectory.
Such beam shaping is achieved by a meta-surface, which diffracts the impinging circular wavefront to be modulated appropriately.
For that purpose, let us consider the diffraction of a monochromatic converging wave by a diffracting optical element.
In this paper, for the sake of clarity, we will consider two-dimensional waves propagating in the $xy$ plane. 
However, a generalization in three dimensions may be carried out straightforwardly.
The polarization of the incident focused field and the subsequently diffracted field will be transverse magnetic (TM), in such a way that the magnetic field with time-domain frequency $\omega$  can be set as $\mathbf{H} (x,y,t) = H(x,y) \exp(-i\omega t) \hat{\mathbf{z}}$,  where $\hat{\mathbf{z}}$ is the unitary vector pointing along the $z$ axis.  
In addition, this will be of help in order to excite surface plasmon resonances in the elementary nano-slits taking part of the designed meta-surface.

\begin{figure}[htbp]
\centerline{\includegraphics[width=0.55\columnwidth]{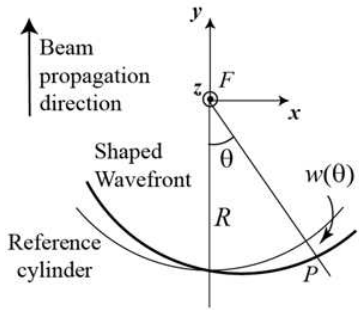}}
\caption{Schematic arrangement representing the diffracted  converging wave of focal point $F$ as evaluated from the Debye  diffraction Eq.~\eqref{eq1}.
 The emerging wave field propagates in the $xy$ plane, and the meta-surface induces a wavefront deviation from its original shape (reference cylinder) by $w(\theta)$. 
 A beam carrying a given acceleration leads to wave localization along an incomplete circumference with center at focus $F$.}
\label{fig01}       
\end{figure}

According  to the Debye diffraction theory, the wave field in the focal region of the shaped converging beam can be obtained by means of the following diffraction integral \cite{Visser92,Zapata07}    
\begin{equation}
H(\mathbf{r})=\sqrt{\frac{k R}{2 \pi i}} e^{i k R} \int \limits_{-\pi/2}^{\pi/2} H_{s}(\theta) e^{\left[ -i k (\mathbf{\hat{q}} \cdot \mathbf{r}) \right]} \mathrm{d} \theta ,
\label{eq1}
\end{equation}
where $\mathbf{r} = (x,y)$, $k = \omega / c$ is the wave number, and $H_{s}$ is the magnetic field over the  reference cylinder. As it is shown in Fig.~\ref{fig01}, $R$  is the radius of the diffracted cylindrical  wave  as  taken  over  the  reference  cylinder. 
Finally, $\mathbf{\hat{q}} = (\sin \theta, \cos \theta)$ is a unit vector pointing from the focus $F$ in the direction of a given point $P$ on the
wavefront.
The scattered wave field $H_{s}(\theta)$ will be expressed by means of a real and positive term, $A(\theta)$, and a phase only term $\exp[i w(\theta)]$, giving
\begin{equation}
H_{s}(\theta) = A(\theta) \exp[i w(\theta)],
\label{eq2}
\end{equation}
The apodization function $A(\theta)$ takes into account the truncation and local attenuation of the converging field after passing through the diffracting optical element, and $w(\theta)$ denotes any deviation of the diffracted wavefront from the reference cylinder, which commonly is interpreted as monochromatic aberration.

The apodization function in our simulations is a super-Gaussian function instead of a customary flat-top profile, which can be written as,
\begin{equation}
A(\theta) = A_{0} \exp[-(\theta / \Omega)^{6}],
\label{eq3}
\end{equation}
of semi-aperture angle $\Omega$, enabling to minimize edge effects. 
We should  mention that the limits of integration in Eq.~\eqref{eq1} including forward-only propagating waves makes the diffracted fields practically unaltered for semi-apertures $\Omega = \pi/2$ and higher values. 
Furthermore we will consider a linear phase modulation, which induces the acceleration of the diffracted field around the focal point $F$ \cite{naserpour2014highly}.
By neglecting a constant term of Taylor expansion of  $w(\theta)$, which has no significant contribution in the diffraction integral Eq.~\eqref{eq1}, and also neglecting higher orders of this expansion, we can rewrite the phase term in Eq.~\eqref{eq2} as $exp(i m \theta)$.
In fact, the parameter $m$ governs the curvilinear trajectory of the focused field around the geometrical focus $F$, as demonstrated below.
Particularly, the distance between the focal point $F$ and where the beam is localized, is determined by $m$. It can be used for controlling the acceleration of a focused beam.
Before the diffraction broadening overcomes, the beam propagates along a circular trajectory whose radius can be estimated as $r_{m}= |m|/k$ \cite{Zapata14c}. 

Analytical solution of the Debye integral Eq.~\eqref{eq1} can be achieved provided that $\Omega \to \infty$ and including backward propagating waves.
By considering the parameter $m$ as an integer number, the magnetic field $H(\mathbf{r})$ finally yields an expression that is proportional to the Bessel function, namely $J_{m}(kr)$, where $r = |\mathbf{r}|$ \cite{naserpour2014highly}. 
Disregarding backward propagation, the resulting wave resembles a sector Bessel field centered on the positive (negative) $x$ axis for $m > 0$ ($m < 0$).  
Because of that, this type of accelerating beams can be called as incomplete Bessel beams \cite{Kaminer12}.
\begin{figure}[htbp]
 \centerline{\includegraphics[width=.8\columnwidth]
 {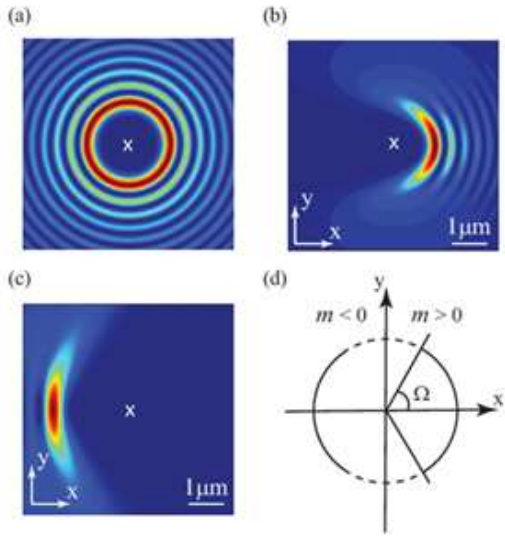}}
 \caption{ (a) Intensity of the magnetic field $H$ corresponding to a Bessel wave field of order $m = 10$ propagating at $\lambda = 632.8\ \mathrm{nm}$. 
 Accelerating wave field with (b) semi-aperture $\Omega = \pi/2$ and $m = + 10$, and (c) $\Omega = \pi/4$ and $m = - 20$. 
 (d) Schematics of the circular trajectory of an incomplete Bessel field.}
 \label{fig02_1} 
\end{figure}
In Fig.~\ref{fig02_1}a and (b) we show the intensity distribution $|H|^2$ of a Bessel field of order $m = 10$ and the associated incomplete Bessel beam of semi-aperture $\Omega = \pi/2$.
By changing the sign of the parameter $m$, the acceleration is reversed however maintaining the center of curvature.
This is illustrated in Fig.~\ref{fig02_1}c and \ref{fig02_1}d.
It is noteworthy
to mention that increasing the absolute value of the order $m$ results in increasing the distance between localized accelerating fields and center of coordinates.

\section{Symmetrization of the incomplete Bessel beam}
\label{sec:2}
\begin{figure}[htbp]
 \centerline{\includegraphics[width=.4\columnwidth]
 {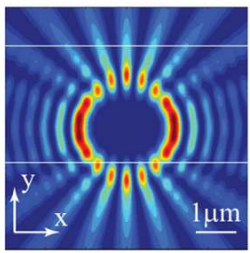}}
 \caption{Superposition of two Bessel-driven accelerating beams of the same semi-aperture $\Omega = \pi/2$ and order $m = 10$ and $m = -10$, respectively.
 Again $\lambda = 632.8\ \mathrm{nm}$.
 An intereference pattern of the field intensity is evident along the $y$ axis.}
 \label{fig02_2} 
\end{figure}

The coherent superposition of accelerating beams with Bessel signatures may be used to create new beam shapes.
For instance, by combining two incomplete Bessel fields of the same order $m$ but opposite sign, we obtain a mirror-symmetric distribution of light, as illustrated in Fig.~\ref{fig02_2}.
In this case, the phase-only term $\exp (i m \theta)$ shown in Eq.~\eqref{eq2} will be substituted by the real-valued periodic modulation $2 \cos(m \theta)$.
Note that an intereference pattern may arise in the vicinities of the $y$ axis for high semi-apertures $\Omega$.

The term $\exp(i m \theta)$ in Eq.~(\ref{eq2}) behaves like a phase term produced by a blazed grating obtaining a maximum diffraction efficiency at a distance $|m|/k$ from focal point.
In fact, a focused beam, associated with an incomplete Bessel wave field of order $m=0$, can be transformed into an accelerating beam by a circular blazed grating.
Furthermore, a meta-surface periodically modulating the amplitude of a focused wave, instead of its phase distribution, will introduce a symmetrization of the accelerated field with respect to the $y$ axis. 
The latter represents the basis of our study. 

\begin{figure}[htbp]
 \centerline{\includegraphics[width=.9\columnwidth]
 {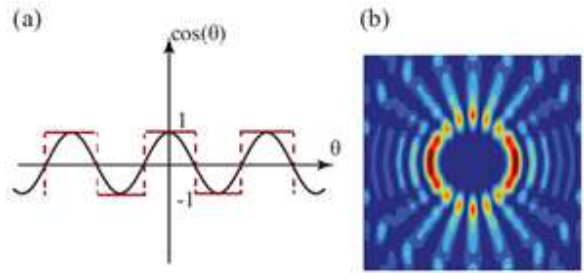}}
 \caption{Binarization of the modulation function $\cos (m \theta)$ and its influence in the Debye diffraction integral:
 (a) schematic of the binary function and (b) resulting field intensity of the diffracted field by using such binary function for $m = 10$ and $\lambda = 632.8\ \mathrm{nm}$.}
 \label{fig03} 
\end{figure}

Here we consider the generation of mirror-symmetric accelerating beams by using diffracting meta-surfaces that shape a given impinging focused wave near its focus.
For practical reasons, the modulation of the converging wave will be carried out more efficiently in phase rather than in amplitude.
Therefore we conveniently binarize the cosine function, as shown in Fig.~\ref{fig03}a.
This transformation will not significantly modify the main contribution to the diffracted field, that is carried by the first diffraction order.
However noisy sidelobes corresponding to high diffraction orders may appear.
Fig.~\ref{fig03}b illustrates the resulting field intensity $|H(\mathbf{r})|^{2}$ whose phase distribution is governed by the binary function of periodicity $2 \pi / m$.
Note that the resultant intensity slightly deviates from that shown in Fig.~\ref{fig02_2}.

\begin{figure}[htbp]
 \centerline{\includegraphics[width=0.9\columnwidth]
 {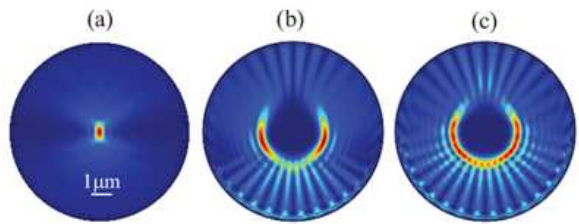}}
 \caption{ Intensity of the magnetic field numerically evaluated when using a surface current (SC).
 In (a) we use a super-Gaussian apodization and in (b) and (c) we additionally modulate the phase distribution of the current by a piecewise binary function of $\pi$ dephase.
 The period of the modulation is $2 \pi / m$, where $m = 16$.}
 \label{fig04} 
\end{figure}

As a proof of concept, in Fig.~\ref{fig04} we show some simple numerical experiments performed by means of a finite-element-analysis commercial software \cite{Comsol}.
A uniform surface current (SC) that is set at a given distance from the origin of coordinates will excite a wave field that focuses at $F$, as shown in Fig.~\ref{fig04}a, where the applied wavelength is $632.8\ \mathrm{nm}$.
By introducing a binary modulation of the surface current of angular period $2 \pi / |m|$, the emitted field propagates near the focus exhibiting a spatial acceleration and a mirror symmetry with respect to the optical axis.
The surface current distribution is additionally apodized by the super-Gaussian distribution $A(\theta)$; we also show the response for a parameter $m = 16$ at different semi-apertures: $\Omega = 3\pi/8$ in Fig.~\ref{fig04}b and $\Omega = \pi/2$ in (c). 
A periodic modulation of dephase $\pi$ produced in adjacent domains will produce a light capsule provided that the semi-aperture $\Omega$ remains close to $\pi/2$, that is in agreement with our analysis based on the Debye diffraction integral.
Nevertheless, even in the optimal case of $\Omega = \pi/2$, the hollow beam is not fully closed due to the proximity of the surface current to the caustic curve.
Otherwise, increasing the radius of the surface current, the field distribution approaches that computed by Eq.~\eqref{eq1}.

\section{Meta-surface design}
\label{sec:3}
\begin{figure}[htbp]
 \centerline{\includegraphics[width=0.9\columnwidth]
 {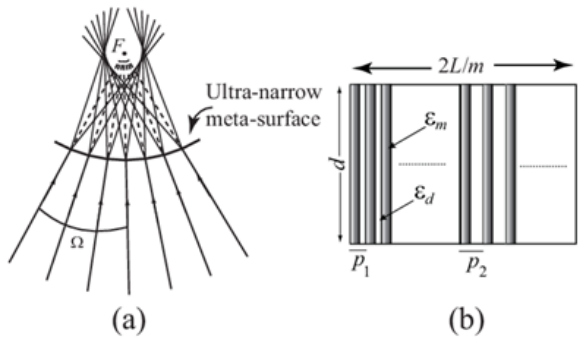}}
 \caption{(a) Geometrical interpretation of the beam shaping using optical rays.
 The impinging rays split and symmetrically bent their trajectories with the same angle.
 The resulting caustic curve is formed by two concentric circular segments of the same radius.
 (b) Design of two periodic media that modulate a transmitted wave inducing a dephase of $\pi$ radians.
 For the sake of clarity we show the basic nanostructured arrangement in a planar geometry.}
 \label{fig05} 
\end{figure}

As shown above, combining two incomplete Bessel beams with positive and negative acceleration parameter $m$ may be achieved by molding the cylindrical wavefront of a converging wave introducing a phase-only binarization.
Such beam shaping will be carried out by a properly-designed plasmonic meta-surface.
A geometrical interpretation of the beam shaping using optical rays is illustrated in Fig.~\ref{fig05}a.
In the simple yet efficient model followed previously, the meta-surface will be segmented into regions wherein individually we induce a nearly-flat phase response in transmission, in a way that additionally neighbouring domains exhibit a dephase of $\pi$.

In order to generate our light capsule we propose a locally-distributed periodic metal-dielectric media. The manipulation of light, such as beam acceleration or focusing in free space, can be achieved by controlling surface plasmon polaritons (SPPs) that are excited at the entrance surface of metallic slits and corrugations with a subwavelength size \cite{Ishii11,Choi12,Tang13}.
For instance, using metallic slit arrays we may control phase retardation between adjacent slits \cite{verslegers2009planar}. 
However, when these nano-slits are tightly arranged, plasmonic modes couple leading to a (certainly limited) degree of homogeneization of the wave fields  \cite{Smith06}.
As a result, it can be used to induce a controlled phase delay.  

The dephase of wave passing through the grating is evaluated by $\varphi = \mathrm{Re}(\beta) d$, where $d$ is the propagation distance in the grating and $\beta$ is the complex propagation constant computed from dispersion
relation of periodic metal-dielectric multilayers medium for TM-polarized waves which is achieved by the Floquet-Bloch theorem.
Considering a multilayered structure of period $\Lambda$ composed of a metal and a dielectric of permittivity $\epsilon_{m}$ and $\epsilon_{d}$, respectively, and using the transfer matrix formulation we finally obtain \cite{Yeh88} 
\begin{equation}\label{Bloch} 
 \cos(K \Lambda) = \cos(\varphi_m) \cos(\varphi_d) - \nu \sin(\varphi_m) \sin(\varphi_d) , 
\end{equation}
where
\begin{equation}
 \nu = \frac{1}{2} \left( \frac{\epsilon_m \kappa_{d}}{\epsilon_d \kappa_{m}} + 
                          \frac{\epsilon_d \kappa_{m}}{\epsilon_m \kappa_{d} } \right).
\end{equation}
In Eq.~\eqref{Bloch}, $\varphi_{q}=\kappa_{q} w_{q}$, $q=\{ m,d \}$ refers to either the metal or the dielectric, 
\begin{equation}
 \kappa_{q} = \sqrt{\epsilon_{q} \kappa^{2}-\beta^{2}} ,
\end{equation}
is the propagation constant in either the metal or the dielectric along the direction that is perpendicular to the metal-dielectric interface, $w_{q}$ denotes the width of each material layer, and $K$ is the Bloch wave number.
Since, in the simulations is considered that the incident wave impinges normally to the metallic grating, Eq.~\eqref{Bloch} is solved for $K=0$ (propagation all along the slits). 
Specifically the metal was gold with $\epsilon_{m}=-10.77+0.79i$ \cite{Lide07} at the wavelength of interest $\lambda=632.8\ \mathrm{nm}$; also the dielectric under consideration was air.
 
According to  Eq.~\eqref{Bloch}, the dephase $\varphi$ gained by a Bloch mode is highly dependent on the width of the layers. 
Therefore, the phase retardation which is required for beam shaping can be obtained by controlling variation of the width of metal slabs and the periodicity of the array.
Finally, an alternating sequence of two dissimilar multilayered metamaterials, displaced transversally to the direction of propagation of the incident field to produce a beam shaping, will give rise to the required binary phase distribution.
This is illustrated in Fig.~\ref{fig05}b.
In our design, the width of metal is kept fixed, $w_{m} = 100\ \mathrm{nm}$, however the grating periods are $\Lambda_{1} = 120\ \mathrm{nm}$ and $\Lambda_{2} = 142\ \mathrm{nm}$. 
The separation between the entrance plane and the exit plane, $d = 600\ \mathrm{nm}$, that is the width of the meta-surface, remains below the wavelength.
As a consequence, the dephase $\mathrm{Re}(\Delta \beta)d$ induced by adjacent photonic crystals will yield $\pi$ radians.
It should be noted that there is certain flexibility in order to attain the required phase distribution \cite{naserpour2014highly}.

Next we will consider a subwavelength meta-surface in such a way that the metallic nanoelements are arranged over a circular sector.
The effective length of the compound meta-surface, that is the part that will be illuminated by the cylindrical wave field, as measured on the exit surface, is equal to $L = 2 \Omega R$, where $R$ is the inner radius of the curvilinear nanostructure.
In our numerical simulations $R = 4 \mu m$ and $\Omega = \pi/2$, and consequently the meta-surface has to include a minimum of $m/2$ bi-grating groups (periods). 
A converging super-Gaussian beam impinges on the circular surface of higher radius, subsequently propagating inside metal-dielectric meta-surface for a distance of $d = 600\ \mathrm{nm}$, where the metal-dielectric surfaces are concentric to the focal point of the incident wave field. 
We should point out that due to the curvilinear distribution of the meta-surface, the layer widths will linearly increase upon distance to the focal point $F$.
However, our approach is still valid since the propagation constant $\beta$ will be practically kept unaltered for a given metal filling fraction \cite{naserpour2014highly}.

\begin{figure}[htbp]
 \centerline{\includegraphics[width=.9\columnwidth]
 {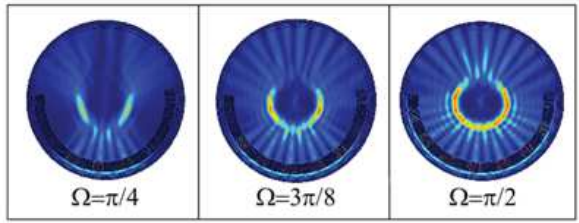}}
 \caption{Intensity distribution of magnetic field for a mirror-symmetric incomplete Bessel beam depending on different angular aperture. 
 With  $\Omega = \pi/2$, a nearly-closed capsule light can be created.}
 \label{fig06} 
\end{figure}

Finally we will show the performance of the designed meta-surface to produce a mirror-symmetric incomplete Bessel beam.
In Fig.~\ref{fig06} we show the intensity of the magnetic field resulting from our finite-element analysis for different angular semi-aperture $\Omega$.
A circular (rather than flat) meta-surface composed of a metallic bi-grating is placed behind a surface current, the former inducing a phase modulation and the latter generating a super-Gaussian aberration-free focused field.
We demonstrate that a nearly-closed capsule light can be created by using a high numerical aperture focused field with $\Omega = \pi/2$.
However, the optical bottle increasingly opens for lower numerical apertures. 

\begin{figure}[htbp]
 \centerline{\includegraphics[width=.9\columnwidth]
 {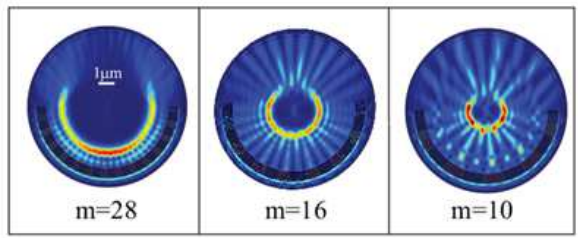}}
 \caption{Intensity distribution of magnetic field for different acceleration parameter $m$, but maintaining $\Omega = \pi/2$. }
 \label{fig07} 
\end{figure}

Analyzing the results shown in Fig.~\ref{fig06}, one can realize that the simulations are in good agreement with theory, as shown in Fig.~\ref{fig04} where the mirror-symmetry incomplete Bessel wave field is derived from an ideal binary phase modulation. 
To conclude we investigate the effect of the acceleration parameters $m$.
Since the radius of Bessel beams is dependent on $m$, we can generate a light capsule with higher radius by simply increasing the value of $m$. 
Fig.~\ref{fig07} shows some numerical simulations for different values of $m$, illustrating the radial tuning of the optical bottle.

\section{Discussion and conclusion}
\label{sec:4}
In conclusion, we proposed a simple method to shape a high-aperture focused field, in the near field, into a nonparaxial accelerating field exhibiting mirror symmetry and that can be used as an optical capsule. 
Through both wave and ray optics we showed how axi-symmetric circular caustics can result from a binary phase modulation.
The resultant nonparaxial hollow fields show a Bessel signature and have  beam-widths near the diffraction limit. 
While these fields are two-dimensional, spherical light shells can be easily produced by using meta-surfaces modulating the wavefront in the azimuthal coordinate. 
Also, the paths followed by the maxima will be circular due to the use of solutions separable in spherical coordinates. 
Our procedure can also be used for producing 1D or 2D optical bottles with arbitrary convex caustics, which is governed by the shape of the diffacting meta-surface, thus providing a versatile tool for all kinds of accelerating waves.

The engineered light capsules are easily sizable and exert forces that can be exploited to realize micrometer-size optical bottles that could attract microparticles or cells.
The beam shaping proposed here is expected to offer certain advantages over the direct generation of accelerating beams using phase masks only, such as the ability to combine mechanical and optical traps produced in the vicinities of the focal region. 
Potential applications also include light-induced curved plasma channels, self-bending electron beams, and accelerating plasmons.

\begin{acknowledgements}
This research was funded by the Spanish Ministry of Economy and Competitiveness under the project TEC2013-50416-EXP. M. Naserpour acknowledges financial support from the Ministry of Science, Research and Technology of Iran.  
\end{acknowledgements}





\end{document}